\documentstyle[12pt]{article}
\begin{document}
\baselineskip 24pt
\noindent
{\large\bf The relations among two transversal submanifolds and global\\
manifold}

\vskip 24pt
\leftskip 9mm
\rightskip 9mm
\normalsize
\baselineskip 24pt
\noindent
Guo-Hong Yang$^{a)}$\footnote{$^{a)}$ E-mail: ghyang@fudan.edu.cn} and
Guang-Jiong Ni\\
{\it Department of Physics, Fudan University, Shanghai, 200433, P.R. China}\\
Yi-Shi Duan\\
{\it Institute of Theoretical Physics, Lanzhou University, Lanzhou, 730000,
P.R. China}

\vskip 24pt
\noindent
In Riemann geometry, the relations among two transversal 
submanifolds and global manifold are discussed. By replacing the normal 
vector of a submanifold with the tangent vector of another submanifold, the 
metric tensors, Christoffel symbols and curvature tensors of the three
manifolds are linked together. When the inner product of the two tangent
vectors vanishes, some corollaries of these relations give the most important
second fundamental form and Gauss-Codazzi equation in the conventional
submanifold theory. As a special case, the global manifold is Euclidean is
considered. It is pointed out that, in order to obtain the nonzero
energy-momentum tensor of matter field in a submanifold, there must be the
contributions of the above inner product and the other submanifold. In general
speaking, a submanifold is closely related to the matter fields of the other
submanifold through the above inner product. This conclusion is in agreement
with the Kaluza-Klein theory and it can be applied to generalize the models of
direct product of manifolds in string and D-brane theories to the more general
cases --- nondirect product manifolds.

\newpage
\leftskip 0mm
\rightskip 0mm

\noindent
{\bf I. INTRODUCTION}

\vskip 24pt
\indent Submanifold theory has reached a fruitful stage. It has been used
extensively in supersymmetry and supergravity$^{1,2}$ and, in particular, in
strings, D-branes and M-theory.$^{3-6}$ In one of our recent papers,$^7$ using
the fourth-order topological tensor current and two transversal submanifolds,
we successfully obtain a new coordinate condition in general relativity, which
includes the Fock's coordinate condition as a special case. Then, with the
help of the Gauss-Bonnet-Chern theorem and the transversal submanifold theory,
we investigate the inner structure of the Euler-Poincar\'{e} characteristic in
differential geometry in Ref. 8, which gives the global, or, topological
relationship of the global manifold and the two transversal submanifolds.
However, in our follow-up studies of the local, or, geometrical relations of
the three manifolds, we find that thought many important formulas (such as the
induced metric tensor, the second fundamental form and the Gauss-Codazzi
equation) are obtained in terms of the tangent and normal vectors of
submanifold, there are still two problems remained in the conventional
submanifold theory of Riemann geometry. (i) The conventional submanifold
theory of Riemann geometry deals with the relationship mainly between the
global manifold and a single submanifold, and it always derives the geometry
of submanifold from the global manifold. Can we reverse the direction, or in
other words, can we build up the geometric quantity of global manifold by the
correspondents of submanifolds? If we can, we will have the relations among
submanifolds and the global manifold, and know how the submanifolds affect
each other when the geometry of the global manifold is given. (ii) For the
lack of the concrete expression of normal vector, some material calculations
can not go forward deeply but be changed into a formal deduction. Can we
replace the normal vector by some other vectors that play the role of normal
vector under some conditions? If the answer is yes, we can extend the
conventional submanifold theory of Riemann geometry. In this paper, we will
discuss these two problems with the two transversal submanifolds. The results
can be obviously used to generalize the models of direct product of manifolds
in string and D-brane theories to the more general cases --- nondirect product
manifolds.\\
\indent This paper is organized as follows. In section II, as a brief review,
we introduce the conventional submanifold theory of Riemann geometry.$^9$ Some
useful notations are also prepared. In section III, we study the relations of
metric tensors, Christoffel symbols and curvature tensors of the two
transversal submanifolds and global manifold. When the inner product of the
tangent vectors of the two transversal submanifolds vanishes, these new
relations give the conventional submanifold theory. As a special case, the
global manifold is Euclidean is considered in section IV. Using the Einstein
equation, the energy-momentum tensor of a submanifold is investigated. The
conclusion of this paper is summarized in section V.

\vskip 48pt
\noindent
{\bf II. THE CONVENTIONAL SUBMANIFOLD THEORY OF RIEMANN GEOMETRY}

\vskip 24pt
\indent Let $X$ be a $k$-dimensional Riemann manifold with metric tensor
$g_{\mu\nu}$ and local coordinates $x^\mu$ $(\mu,\nu=1,\cdots,k)$, and $M$ a
$m$-dimensional submanifold of $X$ with local coordinates $u^a$
$(a=1,\cdots,m;$ $m<k)$. Then, on $M$, one has
\begin{equation}
x^\mu=x^\mu(u^1,\cdots,u^m), \hskip 0.3in \mu=1,\cdots,k.
\end{equation}
The tangent vector basis of $M$ can be expressed in terms of that of $X$ as
\begin{equation}
\frac{\partial}{\partial u^a}=B^\mu_a\frac{\partial}{\partial x^\mu},
\hskip 0.3in B^\mu_a=\frac{\partial x^\mu}{\partial u^a},
\end{equation}
and the induced metric tensor $g_{ab}$ on $M$ is determined by
\begin{equation}
g_{ab}=g_{\mu\nu}B^\mu_aB^\nu_b, \hskip 0.3in a,b=1,\cdots,m.
\end{equation}
With the definition of the normal vector $L^\mu_A$ of $M$
\begin{equation}
g_{\mu\nu}B^\mu_aL^\nu_A=0, \hskip 0.3in
g_{\mu\nu}L^\mu_AL^\nu_B=\delta_{AB}, \hskip 0.3in A,B=m+1,\cdots,k
\end{equation}
and the inverse matrix $(B^a_\mu$, $L^A_\mu)$ of the matrix 
$(B^\mu_a$, $L^\mu_A)$
\begin{equation}
B^a_\mu B^\mu_b=\delta^a_b, \hskip 0.3in L^A_\mu L^\mu_B=\delta^A_B,
\end{equation}
\begin{equation}
B^a_\mu L^\mu_A=0, \hskip 0.3in L^A_\mu B^\mu_a=0,
\end{equation}
\begin{equation}
B^\mu_aB^a_\nu+L^\mu_AL^A_\nu=\delta^\mu_\nu,
\end{equation}
one can prove the completeness relations
\begin{equation}
g_{\mu\nu}=g_{ab}B^a_\mu B^b_\nu+\delta_{AB}L^A_\mu L^B_\nu
\end{equation}
\begin{equation}
g^{\mu\nu}=g^{ab}B^\mu_aB^\nu_b+\delta^{AB}L^\mu_AL^\nu_B
\end{equation}
and the following formulas
\begin{equation}
g_{ab}B^b_\mu=g_{\mu\nu}B^\nu_a, \hskip 0.3in 
\delta_{AB}L^B_\mu=g_{\mu\nu}L^\nu_A,
\end{equation}
\begin{equation}
g^{ab}B^\mu_b=g^{\mu\nu}B^a_\nu, \hskip 0.3in
\delta^{AB}L^\mu_B=g^{\mu\nu}L^A_\nu,
\end{equation}
\begin{equation}
g^{ab}=g^{\mu\nu}B^a_\mu B^b_\nu, \hskip 0.3in
\delta^{AB}=g^{\mu\nu}L^A_\mu L^B_\nu,
\end{equation}
\begin{equation}
g^{\mu\nu}B^a_\mu L^A_\nu=0,
\end{equation}
where $g^{\mu\nu}$ and $g^{ab}$ are the inverses of $g_{\mu\nu}$ and 
$g_{ab}$, respectively. The expressions of $B^a_\mu$ and $L^A_\mu$ can be
obtained simply by raising and lowering the indices $\mu$, $a$ and $A$ of
$B^\mu_a$ and $L^\mu_A$, i.e.
\begin{equation}
B^a_\mu=g^{ab}g_{\mu\nu}B^\nu_b, \hskip 0.3in
L^A_\mu=\delta^{AB}g_{\mu\nu}L^\nu_B.
\end{equation}
After some simple calculations, the relationship between the Christoffel
symbols of $M$ and $X$ can be written as
\begin{equation}
\Gamma^c_{ab}=B^\mu_aB^\nu_bB^c_\lambda\Gamma^\lambda_{\mu\nu}
+B^c_\mu\frac{\partial B^\mu_a}{\partial u^b}.
\end{equation}
The covariant derivative of $B^\mu_a$ is defined by
\begin{equation}
H^\mu_{ab}\equiv\nabla_aB^\mu_b=\frac{\partial B^\mu_b}{\partial u^a}
-\Gamma^c_{ab}B^\mu_c+B^\nu_a\Gamma^\mu_{\nu\lambda}B^\lambda_b
\end{equation}
which is called the Euler-Schouten curvature tensor. Substituting (15) into
(16) and using (7), the Euler-Schouten curvature $H^\mu_{ab}$ is changed into
\begin{equation}
H^\mu_{ab}=(B^\lambda_aB^\rho_b\Gamma^\nu_{\lambda\rho}
+\frac{\partial B^\nu_a}{\partial u^b})L^A_\nu L^\mu_A
\end{equation}
which shows $H^\mu_{ab}$ can be expanded by the normal vector $L^\mu_A$. The
expansion coefficients are defined as the second fundamental form
$\Omega^A_{ab}$, i.e.
\begin{equation}
\Omega^A_{ab}\equiv H^\mu_{ab}L^A_\mu
=(B^\lambda_aB^\rho_b\Gamma^\nu_{\lambda\rho}
+\frac{\partial B^\nu_a}{\partial u^b})L^A_\nu.
\end{equation}
Therefore, the covariant derivative of $B^\mu_a$ can be rewritten as
\begin{equation}
H^\mu_{ab}=\nabla_aB^\mu_b=\Omega^A_{ab}L^\mu_A,
\end{equation}
which leads to
\begin{equation}
H^\mu_{ab}B^c_\mu=B^c_\mu\nabla_aB^\mu_b=\Omega^A_{ab}L^\mu_AB^c_\mu=0.
\end{equation}
Making use of (20) and the generalized Ricci formula
\begin{equation}
R^{\ \ \ d}_{abc}=B^\mu_aB^\nu_bB^\lambda_cB^d_\rho
R^{\ \ \ \ \rho}_{\mu\nu\lambda}-B^d_\mu\nabla_a\nabla_bB^\mu_c
+B^d_\mu\nabla_b\nabla_aB^\mu_c,
\end{equation}
one obtains the most important Gauss-Codazzi equation in the conventional
submanifold theory of Riemann geometry
\begin{equation}
R^{\ \ \ d}_{abc}=B^\mu_aB^\nu_bB^\lambda_cB^d_\rho
R^{\ \ \ \ \rho}_{\mu\nu\lambda}+\nabla_aB^d_\mu\nabla_bB^\mu_c
-\nabla_bB^d_\mu\nabla_aB^\mu_c
\end{equation}
where $R^{\ \ \ d}_{abc}$ and $R^{\ \ \ \ \rho}_{\mu\nu\lambda}$ are the
curvature tensors of $M$ and $X$, respectively.\\
\indent So far, one can see that all of the above consequences are based on
the normal vector $L^\mu_A$, i.e. the definition (4). Though the normal
vector takes great success in the conventional submanifold theory, it can
not give the relationship between two submanifolds. In our recent 
papers,$^{7,8}$ we have obtained two transversal submanifolds and, then, we
can replace the normal vector by the tangent vector of another submanifold.
However, in this case, the orthogonal and orthonormal conditions in (4) are
not held in general.

\vskip 48pt
\noindent
{\bf III. THE RELATIONS AMONG TWO TRANSVERSAL SUBMANIFOLDS AND GLOBAL
MANIFOLD}

\vskip 24pt
\indent Besides of the above-discussed submanifold $M$, let $N$ be another
$n$-dimensional submanifold of $X$ with $n=k-m$ and local coordinates $v^A$
$(A=1,\cdots,n)$. The parametric equation of $N$ is
\begin{equation}
x^\mu=x^\mu(v^1,\cdots,v^n).
\end{equation}
Similarly, the tangent vector basis of $N$ can be expressed as
\begin{equation}
\frac{\partial}{\partial v^A}=C^\mu_A\frac{\partial}{\partial x^\mu},
\hskip 0.3in C^\mu_A=\frac{\partial x^\mu}{\partial v^A},
\end{equation}
and the induced metric tensor $g_{AB}$ is given by
\begin{equation}
g_{AB}=g_{\mu\nu}C^\mu_AC^\nu_B, \hskip 0.3in A,B=1,\cdots,n.
\end{equation}
In the following, in order to investigate the relations among $X$, $M$ and
$N$, we will consider the case that $M$ and $N$ are transversal at a point
$p\in X$. Under this condition, we have
\begin{equation}
T_p(X)=T_p(M)+T_p(N),
\end{equation}
where $T_p(X)$, $T_p(M)$ and $T_p(N)$ are the tangent spaces of $X$, $M$ and
$N$ at $p$, respectively. The expression (26) is to say
\begin{equation}
\frac{\partial}{\partial x^\mu}=B^a_\mu\frac{\partial}{\partial u^a}
+C^A_\mu\frac{\partial}{\partial v^A}, \hskip 0.3in k=m+n,
\end{equation}
in which $B^a_\mu$ and $C^A_\mu$ are the expansion coefficients of
$\partial/\partial x^\mu$ in terms of $\partial/\partial u^a$ and
$\partial/\partial v^A$, respectively. Then, substituting (27) into (2) and
(24), we obtain
\begin{equation}
B^\mu_aB^b_\mu=\delta^b_a, \hskip 0.3in B^\mu_aC^A_\mu=0,
\end{equation}
\begin{equation}
C^\mu_AB^a_\mu=0, \hskip 0.3in C^\mu_AC^B_\mu=\delta^B_A.
\end{equation}
While inserting (2) and (24) into (27) gives
\begin{equation}
B^a_\mu B^\nu_a+C^A_\mu C^\nu_A=\delta^\nu_\mu.
\end{equation}
The formulas (28) --- (30) tell us that the expansion coefficients
$B^a_\mu$ and $C^A_\mu$ are just determined by the inverse matrix of the
matrix $(B^\mu_a$, $C^\mu_A)$.\\
\indent Now, we can discuss the relations among $g_{\mu\nu}$, $g_{ab}$ and
$g_{AB}$. Using 
$g_{\mu\nu}=g_{\lambda\rho}\delta^\lambda_\mu\delta^\rho_\nu$, from (3), 
(25) and (30) we can prove
\begin{equation}
g_{\mu\nu}=g_{ab}B^a_\mu B^b_\nu
+g_{\lambda\rho}B^\lambda_aC^\rho_AB^a_\mu C^A_\nu
+g_{\lambda\rho}C^\lambda_AB^\rho_aC^A_\mu B^a_\nu
+g_{AB}C^A_\mu C^B_\nu.
\end{equation}
Corresponding to the orthogonal relation in (4), the inner product of
$\partial/\partial u^a$ and $\partial/\partial v^A$ is defined as
\begin{equation}
g_{aA}\equiv <\frac{\partial}{\partial u^a}, \frac{\partial}{\partial v^A}>
=g_{\mu\nu}B^\mu_aC^\nu_A
\end{equation}
satisfying
\begin{equation}
g_{aA}=g_{Aa}.
\end{equation}
When $C^\mu_A$ is orthogonal to $B^\mu_a$, we have $g_{aA}=0$. So, (31) can
be further read as
\begin{equation}
g_{\mu\nu}=g_{ab}B^a_\mu B^b_\nu+g_{aA}(B^a_\mu C^A_\nu+C^A_\mu B^a_\nu)
+g_{AB}C^A_\mu C^B_\nu
\end{equation}
which is the generalization of the completeness relation (8). When
$g_{aA}=0$ and $g_{AB}=\delta_{AB}$, (34) goes back to (8). (34) is nothing
but what we seek to show the relations among $g_{\mu\nu}$, $g_{ab}$ and
$g_{AB}$. From (34) we see that, besides of the contributions of $g_{ab}$
and $g_{AB}$, there are still the mixed terms of $g_{aA}$ in $g_{\mu\nu}$,
which will lead to very important new results in our later calculations. The
generalizations of the formulas (10) --- (12) are
\begin{equation}
g_{\mu\nu}B^\nu_a=g_{ab}B^b_\mu+g_{aA}C^A_\mu,
\end{equation}
\begin{equation}
g_{\mu\nu}C^\nu_A=g_{AB}C^B_\mu+g_{aA}B^a_\mu,
\end{equation}
\begin{equation}
g^{\mu\nu}B^a_\nu=g^{ab}B^\mu_b-g^{\mu\nu}g^{ab}g_{bA}C^A_\nu,
\end{equation}
\begin{equation}
g^{\mu\nu}C^A_\nu=g^{AB}C^\mu_B-g^{\mu\nu}g^{AB}g_{aB}B^a_\nu,
\end{equation}
\begin{equation}
g^{ab}=g^{\mu\nu}B^a_\mu B^b_\nu+g^{\mu\nu}g^{ac}g_{cA}C^A_\nu B^b_\mu,
\end{equation}
\begin{equation}
g^{AB}=g^{\mu\nu}C^A_\mu C^B_\nu+g^{\mu\nu}g^{AD}g_{aD}B^a_\nu C^B_\mu,
\end{equation}
which will return to (10) --- (12) when $g_{aA}=0$ and $g_{AB}=\delta_{AB}$,
where $g^{AB}$ is the inverse of $g_{AB}$. Comparing (35) --- (40) with
(10) --- (12), we stress that, besides of the differences between the
conventional terms, the most important are the additional terms of $g_{aA}$
because they provide another way of linking the indices $\mu$ and $a$ or
$\mu$ and $A$. Due to the existence of $g_{aA}$, the expressions of
$B^a_\mu$ and $C^A_\mu$ are no longer as in (14) but become the equations of
$B^a_\mu$ and $C^A_\mu$
\begin{equation}
B^a_\mu=g_{\mu\nu}g^{ab}B^\nu_b-g^{ab}g_{bA}C^A_\mu
\end{equation}
\begin{equation}
C^A_\mu=g_{\mu\nu}g^{AB}C^\nu_B-g^{AB}g_{aB}B^a_\mu.
\end{equation}
Eliminating $C^A_\mu$ in (41) and $B^a_\mu$ in (42), we get
\begin{equation}
(g_{ab}-g_{aA}g^{AB}g_{bB})B^a_\mu=g_{\mu\nu}(B^\nu_b-g_{bA}g^{AB}C^\nu_B)
\end{equation}
\begin{equation}
(g_{AB}-g_{aA}g^{ab}g_{bB})C^A_\mu=g_{\mu\nu}(C^\nu_B-g_{aB}g^{ab}B^\nu_b)
\end{equation}
which can be looked upon as another definitions of $B^a_\mu$ and $C^A_\mu$.
It is obvious that (41) --- (44) can go back to (14) when $g_{aA}=0$ and
$g_{AB}=\delta_{AB}$.\\
\indent In the following, we will study the relations among the Christoffel
symbols of $X$, $M$ and $N$. From (34), the relationship among 
$\Gamma^\lambda_{\mu\nu}$, $\Gamma^c_{ab}$ and $\Gamma^D_{AB}$ can be 
calculated out to be
\begin{eqnarray}
\Gamma^\lambda_{\mu\nu} & = &
B^a_\mu B^b_\nu g^{\lambda\rho}B^d_\rho g_{dc}\Gamma^c_{ab}
+C^A_\mu C^B_\nu g^{\lambda\rho}C^E_\rho g_{ED}\Gamma^D_{AB} \nonumber \\
 & & +B^\lambda_a\frac{\partial B^a_\nu}{\partial x^\mu}
+C^\lambda_A\frac{\partial C^A_\nu}{\partial x^\mu}
+\frac{1}{2}\gamma^\lambda_{\mu\nu}
\end{eqnarray}
where
\begin{eqnarray}
\gamma^\lambda_{\mu\nu} & = &
g^{\lambda\rho}B^a_\rho C^A_\nu\frac{\partial g_{aA}}{\partial x^\mu}
+g^{\lambda\rho}C^A_\rho B^a_\nu\frac{\partial g_{aA}}{\partial x^\mu}
+g^{\lambda\rho}B^a_\rho C^A_\mu\frac{\partial g_{aA}}{\partial x^\nu}
\nonumber \\
 & & +g^{\lambda\rho}C^A_\rho B^a_\mu\frac{\partial g_{aA}}{\partial x^\nu}
-g^{\lambda\rho}B^a_\mu C^A_\nu\frac{\partial g_{aA}}{\partial x^\rho}
-g^{\lambda\rho}C^A_\mu B^a_\nu\frac{\partial g_{aA}}{\partial x^\rho}
\end{eqnarray}
is the contribution of the first-order derivative of $g_{aA}$. The effect of
$g_{aA}$ is included in the factors $g^{\lambda\rho}B^d_\rho g_{dc}$ and
$g^{\lambda\rho}C^E_\rho g_{ED}$. From (45), $\Gamma^c_{ab}$ can be 
expressed in terms of $\Gamma^\lambda_{\mu\nu}$ as
\begin{equation}
\Gamma^c_{ab}
=B^\mu_aB^\nu_bg^{cd}B^\rho_dg_{\rho\lambda}\Gamma^\lambda_{\mu\nu}
+g^{cd}B^\rho_dg_{\rho\lambda}\frac{\partial B^\lambda_b}{\partial u^a}.
\end{equation}
When $g_{aA}=0$, from (41) we have 
$g^{cd}B^\rho_dg_{\rho\lambda}=B^c_\lambda$. Then,
\begin{equation}
\Gamma^c_{ab}=B^\mu_aB^\nu_bB^c_\lambda\Gamma^\lambda_{\mu\nu}
+B^c_\lambda\frac{\partial B^\lambda_b}{\partial u^a}
\end{equation}
which is just the formula given in (15). So, we see that the relationship
between $\Gamma^c_{ab}$ and $\Gamma^\lambda_{\mu\nu}$ in the conventional
submanifold theory is a special case of (47) which is only a corollary of
(45). The similar consequences are held for $\Gamma^D_{AB}$, too. In this
sense, (45) is the total relationship among $\Gamma^\lambda_{\mu\nu}$,
$\Gamma^c_{ab}$ and $\Gamma^D_{AB}$. Since
the effect of $g_{aA}$ has been considered in (45) and (47), the covariant
derivative of $B^\mu_a$ can be still defined by
\begin{equation}
\nabla_aB^\mu_b=\frac{\partial B^\mu_b}{\partial u^a}
-\Gamma^c_{ab}B^\mu_c+B^\nu_a\Gamma^\mu_{\nu\lambda}B^\lambda_b.
\end{equation}
However, under the present condition, from (47) $\nabla_aB^\mu_b$ can be
represented by
\begin{equation}
\nabla_aB^\mu_b=(B^\lambda_aB^\rho_b\Gamma^\nu_{\lambda\rho}
+\frac{\partial B^\nu_b}{\partial u^a})C^A_\nu C^\mu_A
-(B^\lambda_aB^\rho_b\Gamma^\nu_{\lambda\rho}
+\frac{\partial B^\nu_b}{\partial u^a})C^A_\nu g_{dA}g^{cd}B^\mu_c
\end{equation}
which shows $\nabla_aB^\mu_b$ can not be expanded by $C^\mu_A$ only
but by $B^\mu_a$ and $C^\mu_A$ together. This extension is also due to the
existence of $g_{aA}$. Corresponding to the second fundamental form defined
in (18), let us denote the expansion coefficients as
\begin{equation}
\Theta^A_{ab}\equiv (\nabla_aB^\mu_b)C^A_\mu, \hskip 0.3in
\Phi^c_{ab}\equiv (\nabla_aB^\mu_b)B^c_\mu.
\end{equation}
Then,
\begin{equation}
\Theta^A_{ab}=(B^\lambda_aB^\rho_b\Gamma^\nu_{\lambda\rho}
+\frac{\partial B^\nu_b}{\partial u^a})C^A_\nu
\end{equation}
\begin{equation}
\Phi^c_{ab}=-(B^\lambda_aB^\rho_b\Gamma^\nu_{\lambda\rho}
+\frac{\partial B^\nu_b}{\partial u^a})C^A_\nu g_{dA}g^{dc}
\end{equation}
and
\begin{equation}
\nabla_aB^\mu_b=\Theta^A_{ab}C^\mu_A+\Phi^c_{ab}B^\mu_c.
\end{equation}
One may have noticed that the expansion coefficients are not independent.
They are related by
\begin{equation}
\Phi^c_{ab}=-\Theta^A_{ab}g_{dA}g^{dc}.
\end{equation}
Though $\nabla_aB^\mu_b$ can not
be expanded by $C^\mu_A$ only, we find $\nabla_a(g^{bc}g_{\mu\nu}B^\nu_c)$
does, i.e.
\begin{eqnarray}
\nabla_a(g^{bc}g_{\mu\nu}B^\nu_c) & = &
\frac{\partial(g^{bc}g_{\mu\nu}B^\nu_c)}{\partial u^a}
+\Gamma^b_{ac}g^{cd}g_{\mu\nu}B^\nu_d
-B^\nu_a\Gamma^\lambda_{\nu\mu}g^{bc}g_{\lambda\rho}B^\rho_c \nonumber  \\
 & = & \frac{\partial(g^{bc}g_{\mu\nu}B^\nu_c)}{\partial u^a}
-B^\nu_a\Gamma^\lambda_{\nu\mu}g^{bc}g_{\lambda\rho}B^\rho_c
+\Gamma^b_{ac}g^{cd}g_{dA}C^A_\mu+\Gamma^b_{ac}B^c_\mu \nonumber \\
 & = & (\frac{\partial(g^{bc}g_{\nu\lambda}B^\lambda_c)}{\partial u^a}
+\Gamma^b_{ac}g^{cd}g_{\nu\lambda}B^\lambda_d
-B^\lambda_a\Gamma^\rho_{\lambda\nu}g^{bc}g_{\rho\alpha}B^\alpha_c)C^\nu_AC^A_\mu \nonumber \\
 & = & (\nabla_a(g^{bc}g_{\nu\lambda}B^\lambda_c))C^\nu_AC^A_\mu
\end{eqnarray} 
which leads to
\begin{equation}
(\nabla_a(g^{bc}g_{\mu\nu}B^\nu_c))B^\mu_d=0.
\end{equation}
So, $\nabla_a(g^{bc}g_{\mu\nu}B^\nu_c)$ will play the role of 
$\nabla_aB^b_\mu$ in some places, such as in the Gauss-Codazzi equation.\\
\indent Now, let us consider the relations of the curvature tensors of $X$,
$M$ and $N$. Similar to the covariant derivatives of $B^\mu_a$ and $C^\mu_A$
in the conventional submanifold theory, we denote the notations
$\nabla_\mu B^\nu_a$, $\nabla_\mu B^a_\nu$, $\nabla_\mu C^\nu_A$ and
$\nabla_\mu C^A_\nu$ as
\begin{equation}
\nabla_\mu B^\nu_a=B^b_\mu\frac{\partial B^\nu_a}{\partial u^b}
-B^b_\mu\Gamma^c_{ba}B^\nu_c+\Gamma^\nu_{\mu\lambda}B^\lambda_a
\end{equation}
\begin{equation}
\nabla_\mu B^a_\nu=\frac{\partial B^a_\nu}{\partial x^\mu}
+B^b_\mu\Gamma^a_{bc}B^c_\nu-\Gamma^\lambda_{\mu\nu}B^a_\lambda
\end{equation}
\begin{equation}
\nabla_\mu C^\nu_A=C^B_\mu\frac{\partial C^\nu_A}{\partial v^B}
-C^B_\mu\Gamma^D_{BA}C^\nu_D+\Gamma^\nu_{\mu\lambda}C^\lambda_A
\end{equation}
\begin{equation}
\nabla_\mu C^A_\nu=\frac{\partial C^A_\nu}{\partial x^\mu}
+C^B_\mu\Gamma^A_{BD}C^D_\nu-\Gamma^\lambda_{\mu\nu}C^A_\lambda.
\end{equation}
Then, after long and complicated calculations, the curvature tensor
$R^{\ \ \ \ \rho}_{\mu\nu\lambda}$ of $X$ can be expressed by
\begin{eqnarray}
R^{\ \ \ \ \rho}_{\mu\nu\lambda} & = &
B^a_\mu B^b_\nu B^c_\lambda g^{\rho\sigma}B^d_\sigma g_{de}R^{\ \ \ e}_{abc}
+C^A_\mu C^B_\nu C^D_\lambda g^{\rho\sigma}C^E_\sigma g_{EF}
R^{\ \ \ \ \ F}_{ABD} \nonumber \\
 & & +\nabla_\mu B^\rho_a\nabla_\nu B^a_\lambda
+\nabla_\mu C^\rho_A\nabla_\nu C^A_\lambda
-\nabla_\nu B^\rho_a\nabla_\mu B^a_\lambda
-\nabla_\nu C^\rho_A\nabla_\mu C^A_\lambda \nonumber \\
 & & +\Theta+\Lambda+\Phi+\Psi+\Omega+\Sigma
\end{eqnarray}
where $R^{\ \ \ d}_{abc}$ and $R^{\ \ \ \ \ E}_{ABD}$ are the curvature
tensors of $M$ and $N$, and
\begin{eqnarray}
\Theta & = &
-g_{aA}(\frac{\partial B^d_\alpha}{\partial x^\mu}B^b_\nu B^c_\lambda
B^\rho_dg^{\alpha\beta}C^A_\beta\Gamma^a_{bc}
+\frac{\partial C^B_\alpha}{\partial x^\mu}B^b_\nu B^c_\lambda C^\rho_B
g^{\alpha\beta}C^A_\beta\Gamma^a_{bc}
+\frac{\partial(g^{\rho\sigma}C^A_\sigma)}{\partial x^\mu}B^b_\nu
B^c_\lambda\Gamma^a_{bc} \nonumber \\
 & & +\frac{\partial C^E_\alpha}{\partial x^\mu}C^B_\nu C^D_\lambda C^\rho_E
g^{\alpha\beta}B^a_\beta\Gamma^A_{BD}
+\frac{\partial B^b_\alpha}{\partial x^\mu}C^B_\nu C^D_\lambda B^\rho_b
g^{\alpha\beta}B^a_\beta\Gamma^A_{BD} \nonumber \\
 & & +\frac{\partial(g^{\rho\sigma}B^a_\sigma)}{\partial x^\mu}C^B_\nu
C^D_\lambda\Gamma^A_{BD}
+C^B_\mu C^D_\nu C^E_\lambda g^{\rho\sigma}C^F_\sigma g_{FG}C^H_\alpha
g^{\alpha\beta}B^a_\beta\Gamma^G_{BH}\Gamma^A_{DE} \nonumber \\
 & & +C^B_\mu B^b_\nu B^c_\lambda g^{\rho\sigma}C^D_\sigma g_{DE}C^F_\alpha
g^{\alpha\beta}C^A_\beta\Gamma^E_{BF}\Gamma^a_{bc}
+B^b_\mu B^c_\nu B^d_\lambda g^{\rho\sigma}B^e_\sigma g_{ef}B^g_\alpha
g^{\alpha\beta}C^A_\beta\Gamma^f_{bg}\Gamma^a_{cd} \nonumber \\
 & & +B^b_\mu C^B_\nu C^D_\lambda g^{\rho\sigma}B^c_\sigma g_{cd}B^e_\alpha
g^{\alpha\beta}B^a_\beta\Gamma^d_{be}\Gamma^A_{BD}
-\frac{\partial B^d_\alpha}{\partial x^\nu}B^b_\mu B^c_\lambda
B^\rho_dg^{\alpha\beta}C^A_\beta\Gamma^a_{bc} \nonumber \\
 & & -\frac{\partial C^B_\alpha}{\partial x^\nu}B^b_\mu B^c_\lambda C^\rho_B
g^{\alpha\beta}C^A_\beta\Gamma^a_{bc}
-\frac{\partial(g^{\rho\sigma}C^A_\sigma)}{\partial x^\nu}B^b_\mu
B^c_\lambda\Gamma^a_{bc}
-\frac{\partial C^E_\alpha}{\partial x^\nu}C^B_\mu C^D_\lambda C^\rho_E
g^{\alpha\beta}B^a_\beta\Gamma^A_{BD} \nonumber \\
 & & -\frac{\partial B^b_\alpha}{\partial x^\nu}C^B_\mu C^D_\lambda B^\rho_b
g^{\alpha\beta}B^a_\beta\Gamma^A_{BD}
-C^B_\nu C^D_\mu C^E_\lambda g^{\rho\sigma}C^F_\sigma g_{FG}C^H_\alpha
g^{\alpha\beta}B^a_\beta\Gamma^G_{BH}\Gamma^A_{DE} \nonumber \\
 & & -C^B_\nu B^b_\mu B^c_\lambda g^{\rho\sigma}C^D_\sigma g_{DE}C^F_\alpha
g^{\alpha\beta}C^A_\beta\Gamma^E_{BF}\Gamma^a_{bc}
-B^b_\nu B^c_\mu B^d_\lambda g^{\rho\sigma}B^e_\sigma g_{ef}B^g_\alpha
g^{\alpha\beta}C^A_\beta\Gamma^f_{bg}\Gamma^a_{cd} \nonumber \\
 & & -\frac{\partial(g^{\rho\sigma}B^a_\sigma)}{\partial x^\nu}C^B_\mu
C^D_\lambda\Gamma^A_{BD}
-B^b_\nu C^B_\mu C^D_\lambda g^{\rho\sigma}B^c_\sigma g_{cd}B^e_\alpha
g^{\alpha\beta}B^a_\beta\Gamma^d_{be}\Gamma^A_{BD})
\end{eqnarray}
\begin{eqnarray}
\Lambda & = &
g_{aA}g_{bB}(B^c_\mu B^d_\nu B^e_\lambda g^{\rho\sigma}C^B_\sigma B^f_\alpha
g^{\alpha\beta}C^A_\beta\Gamma^a_{de}\Gamma^b_{cf}
+B^c_\mu C^D_\nu C^E_\lambda g^{\rho\sigma}C^B_\sigma B^d_\alpha
g^{\alpha\beta}B^a_\beta\Gamma^b_{cd}\Gamma^A_{DE} \nonumber \\
 & & +C^D_\mu C^E_\nu C^F_\lambda g^{\rho\sigma}B^b_\sigma C^G_\alpha
g^{\alpha\beta}B^a_\beta\Gamma^A_{EF}\Gamma^B_{DG}
+C^D_\mu B^c_\nu B^d_\lambda g^{\rho\sigma}B^b_\sigma C^E_\alpha
g^{\alpha\beta}C^A_\beta\Gamma^a_{cd}\Gamma^B_{DE} \nonumber \\
 & & -B^c_\nu B^d_\mu B^e_\lambda g^{\rho\sigma}C^B_\sigma B^f_\alpha
g^{\alpha\beta}C^A_\beta\Gamma^a_{de}\Gamma^b_{cf}
-B^c_\nu C^D_\mu C^E_\lambda g^{\rho\sigma}C^B_\sigma B^d_\alpha
g^{\alpha\beta}B^a_\beta\Gamma^b_{cd}\Gamma^A_{DE} \nonumber \\
 & & -C^D_\nu C^E_\mu C^F_\lambda g^{\rho\sigma}B^b_\sigma C^G_\alpha
g^{\alpha\beta}B^a_\beta\Gamma^A_{EF}\Gamma^B_{DG}
-C^D_\nu B^c_\mu B^d_\lambda g^{\rho\sigma}B^b_\sigma C^E_\alpha
g^{\alpha\beta}C^A_\beta\Gamma^a_{cd}\Gamma^B_{DE})
\end{eqnarray}
\begin{eqnarray}
\Phi & = &
-\frac{\partial g_{aA}}{\partial x^\mu}B^b_\nu B^c_\lambda g^{\rho\sigma}
C^A_\sigma\Gamma^a_{bc}
-\frac{\partial g_{aA}}{\partial x^\mu}C^B_\nu C^D_\lambda g^{\rho\sigma}
B^a_\sigma\Gamma^A_{BD}
+\frac{1}{2}\frac{\partial g_{aA}}{\partial x^\nu}
\frac{\partial(g^{\rho\sigma}B^a_\sigma C^A_\lambda)}{\partial x^\mu}
\nonumber \\
 & & +\frac{1}{2}\frac{\partial g_{aA}}{\partial x^\nu}
\frac{\partial(g^{\rho\sigma}C^A_\sigma B^a_\lambda)}{\partial x^\mu}
+\frac{1}{2}\frac{\partial g_{aA}}{\partial x^\lambda}
\frac{\partial(g^{\rho\sigma}B^a_\sigma C^A_\nu)}{\partial x^\mu}
+\frac{1}{2}\frac{\partial g_{aA}}{\partial x^\lambda}
\frac{\partial(g^{\rho\sigma}C^A_\sigma B^a_\nu)}{\partial x^\mu}
\nonumber \\
 & & -\frac{1}{2}\frac{\partial g_{aA}}{\partial x^\sigma}
\frac{\partial(g^{\rho\sigma}B^a_\nu C^A_\lambda)}{\partial x^\mu}
-\frac{1}{2}\frac{\partial g_{aA}}{\partial x^\sigma}
\frac{\partial(g^{\rho\sigma}C^A_\nu B^a_\lambda)}{\partial x^\mu}
+\frac{1}{2}B^a_\mu g^{\rho\sigma}B^c_\sigma g_{cd}B^b_\alpha\Gamma^d_{ab}
\gamma^\alpha_{\nu\lambda} \nonumber \\
 & & +\frac{1}{2}C^A_\mu g^{\rho\sigma}C^D_\sigma g_{DE}C^B_\alpha
\Gamma^E_{AB}\gamma^\alpha_{\nu\lambda}
+\frac{1}{2}B^\rho_a\frac{\partial B^a_\alpha}{\partial x^\mu}
\gamma^\alpha_{\nu\lambda}
+\frac{1}{2}C^\rho_A\frac{\partial C^A_\alpha}{\partial x^\mu}
\gamma^\alpha_{\nu\lambda} \nonumber \\
 & & +\frac{1}{2}B^a_\nu B^b_\lambda g^{\alpha\beta}B^c_\beta g_{cd}
\Gamma^d_{ab}\gamma^\rho_{\mu\alpha}
+\frac{1}{2}C^A_\nu C^B_\lambda g^{\alpha\beta}C^D_\beta g_{DE}
\Gamma^E_{AB}\gamma^\rho_{\mu\alpha}
+\frac{1}{2}B^\alpha_a\frac{\partial B^a_\lambda}{\partial x^\nu}
\gamma^\rho_{\mu\alpha} \nonumber \\
 & & +\frac{1}{2}C^\alpha_A\frac{\partial C^A_\lambda}{\partial x^\nu}
\gamma^\rho_{\mu\alpha}
+\frac{\partial g_{aA}}{\partial x^\nu}B^b_\mu B^c_\lambda
g^{\rho\sigma}C^A_\sigma\Gamma^a_{bc}
+\frac{\partial g_{aA}}{\partial x^\nu}C^B_\mu C^D_\lambda g^{\rho\sigma}
B^a_\sigma\Gamma^A_{BD} \nonumber \\
 & & -\frac{1}{2}\frac{\partial g_{aA}}{\partial x^\mu}
\frac{\partial(g^{\rho\sigma}B^a_\sigma C^A_\lambda)}{\partial x^\nu}
-\frac{1}{2}\frac{\partial g_{aA}}{\partial x^\mu}
\frac{\partial(g^{\rho\sigma}C^A_\sigma B^a_\lambda)}{\partial x^\nu}
-\frac{1}{2}\frac{\partial g_{aA}}{\partial x^\lambda}
\frac{\partial(g^{\rho\sigma}B^a_\sigma C^A_\mu)}{\partial x^\nu}
\nonumber \\
 & & -\frac{1}{2}\frac{\partial g_{aA}}{\partial x^\lambda}
\frac{\partial(g^{\rho\sigma}C^A_\sigma B^a_\mu)}{\partial x^\nu}
+\frac{1}{2}\frac{\partial g_{aA}}{\partial x^\sigma}
\frac{\partial(g^{\rho\sigma}B^a_\mu C^A_\lambda)}{\partial x^\nu}
+\frac{1}{2}\frac{\partial g_{aA}}{\partial x^\sigma}
\frac{\partial(g^{\rho\sigma}C^A_\mu B^a_\lambda)}{\partial x^\nu}
\nonumber \\
 & & -\frac{1}{2}B^a_\nu g^{\rho\sigma}B^c_\sigma g_{cd}B^b_\alpha
\Gamma^d_{ab}\gamma^\alpha_{\mu\lambda}
-\frac{1}{2}C^A_\nu g^{\rho\sigma}C^D_\sigma g_{DE}C^B_\alpha
\Gamma^E_{AB}\gamma^\alpha_{\mu\lambda}
-\frac{1}{2}B^\rho_a\frac{\partial B^a_\alpha}{\partial x^\nu}
\gamma^\alpha_{\mu\lambda} \nonumber \\
 & & -\frac{1}{2}C^\rho_A\frac{\partial C^A_\alpha}{\partial x^\nu}
\gamma^\alpha_{\mu\lambda}
-\frac{1}{2}B^a_\mu B^b_\lambda g^{\alpha\beta}B^c_\beta g_{cd}\Gamma^d_{ab}
\gamma^\rho_{\nu\alpha}
-\frac{1}{2}C^A_\mu C^B_\lambda g^{\alpha\beta}C^D_\beta g_{DE}
\Gamma^E_{AB}\gamma^\rho_{\nu\alpha} \nonumber \\
 & & -\frac{1}{2}B^\alpha_a\frac{\partial B^a_\lambda}{\partial x^\mu}
\gamma^\rho_{\nu\alpha}
-\frac{1}{2}C^\alpha_A\frac{\partial C^A_\lambda}{\partial x^\mu}
\gamma^\rho_{\nu\alpha}
\end{eqnarray}
\begin{eqnarray}
\Psi & = &
-\frac{1}{2}g_{aA}(B^b_\mu g^{\rho\sigma}C^A_\sigma B^c_\alpha\Gamma^a_{bc}
\gamma^\alpha_{\nu\lambda}
+C^B_\mu g^{\rho\sigma}B^a_\sigma C^D_\alpha\Gamma^A_{BD}
\gamma^\alpha_{\nu\lambda}
+B^b_\nu B^c_\lambda g^{\alpha\beta}C^A_\beta\Gamma^a_{bc}
\gamma^\rho_{\mu\alpha} \nonumber \\
 & & +C^B_\nu C^D_\lambda g^{\alpha\beta}B^a_\beta\Gamma^A_{BD}
\gamma^\rho_{\mu\alpha}
-B^b_\nu g^{\rho\sigma}C^A_\sigma B^c_\alpha\Gamma^a_{bc}
\gamma^\alpha_{\mu\lambda}
-C^B_\nu g^{\rho\sigma}B^a_\sigma C^D_\alpha\Gamma^A_{BD}
\gamma^\alpha_{\mu\lambda} \nonumber \\
 & & -B^b_\mu B^c_\lambda g^{\alpha\beta}C^A_\beta\Gamma^a_{bc}
\gamma^\rho_{\nu\alpha}
-C^B_\mu C^D_\lambda g^{\alpha\beta}B^a_\beta\Gamma^A_{BD}
\gamma^\rho_{\nu\alpha})
\end{eqnarray}
\begin{equation}
\Omega=\frac{1}{2}\gamma^\rho_{\mu\alpha}\gamma^\alpha_{\nu\lambda}
-\frac{1}{2}\gamma^\rho_{\nu\alpha}\gamma^\alpha_{\mu\lambda}
\end{equation}
\begin{eqnarray}
\Sigma & = &
\frac{1}{2}\frac{\partial^2g_{aA}}{\partial x^\mu\partial x^\lambda}
g^{\rho\sigma}B^a_\sigma C^A_\nu
+\frac{1}{2}\frac{\partial^2g_{aA}}{\partial x^\mu\partial x^\lambda}
g^{\rho\sigma}C^A_\sigma B^a_\nu
-\frac{1}{2}\frac{\partial^2g_{aA}}{\partial x^\mu\partial x^\sigma}
g^{\rho\sigma}B^a_\nu C^A_\lambda \nonumber \\
 & & -\frac{1}{2}\frac{\partial^2g_{aA}}{\partial x^\mu\partial x^\sigma}
g^{\rho\sigma}C^A_\nu B^a_\lambda
-\frac{1}{2}\frac{\partial^2g_{aA}}{\partial x^\nu\partial x^\lambda}
g^{\rho\sigma}B^a_\sigma C^A_\mu
-\frac{1}{2}\frac{\partial^2g_{aA}}{\partial x^\nu\partial x^\lambda}
g^{\rho\sigma}C^A_\sigma B^a_\mu \nonumber \\
 & & +\frac{1}{2}\frac{\partial^2g_{aA}}{\partial x^\nu\partial x^\sigma}
g^{\rho\sigma}B^a_\mu C^A_\lambda
+\frac{1}{2}\frac{\partial^2g_{aA}}{\partial x^\nu\partial x^\sigma}
g^{\rho\sigma}C^A_\mu B^a_\lambda,
\end{eqnarray}
in which $\gamma^\lambda_{\mu\nu}$ includes the first-order partial
derivatives of $g_{aA}$ and is defined in (46). Hence, $\Theta$, $\Lambda$,
$\Phi$, $\Psi$, $\Omega$ and $\Sigma$ are the contributions of $g_{aA}$,
$g_{aA}g_{bB}$, $\partial g_{aA}$, $g_{aA}\partial g_{bB}$,
$\partial g_{aA}\partial g_{bB}$ and $\partial^2g_{aA}$, respectively.
Though the expressions of $R^{\ \ \ \ \rho}_{\mu\nu\lambda}$ in (62) --- (68)
are very complicated and strongly nonlinear, the curvature tensor
$R^{\ \ \ d}_{abc}$ of $M$ can be represented in terms of
$R^{\ \ \ \ \rho}_{\mu\nu\lambda}$ simply as
\begin{eqnarray}
R^{\ \ \ d}_{abc} & = &
B^\mu_aB^\nu_bB^\lambda_cg^{de}B^\sigma_eg_{\sigma\rho}
R^{\ \ \ \ \rho}_{\mu\nu\lambda}
+\nabla_a(g^{de}B^\mu_eg_{\mu\nu})\nabla_bB^\nu_c \nonumber \\
 & & -\nabla_b(g^{de}B^\mu_eg_{\mu\nu})\nabla_aB^\nu_c
\end{eqnarray}
where
\begin{equation}
\nabla_a(g^{de}B^\mu_eg_{\mu\nu})
=\frac{\partial(g^{de}B^\mu_eg_{\mu\nu})}{\partial u^a}
+\Gamma^d_{ab}g^{be}B^\mu_eg_{\mu\nu}
-B^\lambda_a\Gamma^\rho_{\lambda\nu}g^{de}B^\mu_eg_{\mu\rho}
\end{equation}
\begin{equation}
\nabla_bB^\nu_c=\frac{\partial B^\nu_c}{\partial u^b}-\Gamma^a_{bc}B^\nu_a
+B^\mu_b\Gamma^\nu_{\mu\lambda}B^\lambda_c
\end{equation}
are the covariant derivatives of $g^{de}B^\mu_eg_{\mu\nu}$ and $B^\nu_c$.
When $g_{aA}=0$, the two tangent vectors $B^\mu_a$ and $C^\mu_A$ of $M$ and
$N$ are orthogonal each other and $C^\mu_A$ plays the role of the normal
vector of $B^\mu_a$ in the conventional submanifold theory. In this case,
from (41) we have $g^{de}B^\sigma_eg_{\sigma\rho}=B^d_\rho$ and then
\begin{equation}
R^{\ \ \ d}_{abc}=B^\mu_aB^\nu_bB^\lambda_cB^d_\rho
R^{\ \ \ \ \rho}_{\mu\nu\lambda}+\nabla_aB^d_\mu\nabla_bB^\mu_c
-\nabla_bB^d_\mu\nabla_aB^\mu_c
\end{equation}
which is just the most important Gauss-Codazzi equation in the conventional
submanifold theory of Riemann geometry. So, we see that the
Gauss-Codazzi equation is a special case of (69) which is only a corollary
of (62). The similar consequences are held for $R^{\ \ \ \ \ E}_{ABD}$, too.
In this sense, (62) --- (68) are the total relationship among
$R^{\ \ \ \ \rho}_{\mu\nu\lambda}$, $R^{\ \ \ d}_{abc}$ and
$R^{\ \ \ \ \ E}_{ABD}$. Comparing (69) with (72), as indicated in above,
$\nabla_a(g^{de}B^\mu_eg_{\mu\nu})$ takes the place of
$\nabla_aB^d_\nu$ in the general case. The formulas derived in this section
can be applied to generalize the models of direct product of manifolds in
string and D-brane theories to the more general cases --- nondirect product
manifolds.

\vskip 48pt
\noindent
{\bf IV. A SPECIAL CASE}

\vskip 24pt
\indent In this section, we do not pay more attention to the condition
$g_{aA}=0$ which has been studied extensively in Riemann geometry, but change
to consider the case $g_{\mu\nu}=\delta_{\mu\nu}$, i.e. the global manifold
$X$ is Euclidean. In this case, we have
\begin{equation}
g_{\mu\nu}=\delta_{\mu\nu}, \hskip 0.3in \Gamma^\lambda_{\mu\nu}=0,
\hskip 0.3in R^{\ \ \ \ \rho}_{\mu\nu\lambda}=0.
\end{equation}
This is the condition that we obtain a new coordinate condition in general
relativity,$^7$ which includes the Fock's coordinate condition as a special
case. Then, in order to maintain the Euclidean property of $X$, the induced
metric tensors and the Christoffel symbols of $M$ and $N$ must satisfy
\begin{equation}
\delta_{\mu\nu}=g_{ab}B^a_\mu B^b_\nu
+g_{aA}(B^a_\mu C^A_\nu+C^A_\mu B^a_\nu)+g_{AB}C^A_\mu C^B_\nu
\end{equation}
and
\begin{equation}
0=B^a_\mu B^b_\nu\delta^{\lambda\rho}B^d_\rho g_{dc}\Gamma^c_{ab}
+C^A_\mu C^B_\nu\delta^{\lambda\rho}C^E_\rho g_{ED}\Gamma^D_{AB}
+B^\lambda_a\frac{\partial B^a_\nu}{\partial x^\mu}
+C^\lambda_A\frac{\partial C^A_\nu}{\partial x^\mu}
+\frac{1}{2}\gamma^\lambda_{\mu\nu},
\end{equation}
respectively. In this case, $g_{ab}$ and $\Gamma^c_{ab}$ are represented by
\begin{equation}
g_{ab}=\delta_{\mu\nu}B^\mu_aB^\nu_b
\end{equation}
\begin{equation}
\Gamma^c_{ab}=g^{cd}B^\mu_d\delta_{\mu\nu}
\frac{\partial B^\nu_b}{\partial u^a}.
\end{equation}
Here, we point out that (76) is not the vielbein expression of $g_{ab}$
because the dimension of $\mu$ is not equal to that of $a$ and 
$B^\mu_a$ is not the vielbein field. And (77) is quite different from the
corresponding form of $\Gamma^c_{ab}$ in (15) because (77) includes the
contribution of $C^A_\mu$ in $g^{cd}B^\mu_d\delta_{\mu\nu}$ by taking account
of (41). In fact, this conclusion is also held for (47) in the general
case because these two transversal submanifolds must maintain the geometric
property of the given global manifold. On the curvature tensors
$R^{\ \ \ d}_{abc}$ and $R^{\ \ \ \ \ E}_{ABD}$ of $M$ and $N$, we have
\begin{eqnarray}
0 & = &
B^a_\mu B^b_\nu B^c_\lambda\delta^{\rho\sigma}B^d_\sigma g_{de}
R^{\ \ \ e}_{abc}
+C^A_\mu C^B_\nu C^D_\lambda\delta^{\rho\sigma}C^E_\sigma g_{EF}
R^{\ \ \ \ \ F}_{ABD} \nonumber \\
 & & +D_\mu B^\rho_aD_\nu B^a_\lambda+D_\mu C^\rho_AD_\nu C^A_\lambda
-D_\nu B^\rho_aD_\mu B^a_\lambda-D_\nu C^\rho_AD_\mu C^A_\lambda
\nonumber \\
 & & +\Theta'+\Lambda'+\Phi'
\end{eqnarray}
where
\begin{equation}
D_\mu B^\rho_a=B^b_\mu\frac{\partial B^\rho_a}{\partial u^b}
-B^b_\mu\Gamma^c_{ba}B^\rho_c, \hskip 0.3in
D_\nu B^a_\lambda=\frac{\partial B^a_\lambda}{\partial x^\nu}
+B^b_\nu\Gamma^a_{bc}B^c_\lambda
\end{equation}
\begin{equation}
D_\mu C^\rho_A=C^B_\mu\frac{\partial C^\rho_A}{\partial v^B}
-C^B_\mu\Gamma^D_{BA}C^\rho_D, \hskip 0.3in
D_\nu C^A_\lambda=\frac{\partial C^A_\lambda}{\partial x^\nu}
+C^B_\nu\Gamma^A_{BD}C^D_\lambda
\end{equation}
and
\begin{eqnarray}
\Theta' & = &
g_{aA}(B^b_\mu B^d_\nu B^e_\lambda\delta^{\rho\sigma}C^A_\sigma\Gamma^a_{bc}
\Gamma^c_{de}
+C^B_\mu C^E_\nu C^F_\lambda\delta^{\rho\sigma}B^a_\sigma\Gamma^A_{BD}
\Gamma^D_{EF}
+B^b_\mu\delta^{\rho\sigma}C^A_\sigma
\frac{\partial B^c_\lambda}{\partial x^\nu}\Gamma^a_{bc} \nonumber \\
 & & +C^B_\mu\delta^{\rho\sigma}B^a_\sigma\Gamma^A_{BD}
\frac{\partial C^D_\lambda}{\partial x^\nu}
-\frac{\partial C^A_\sigma}{\partial x^\mu}B^b_\nu B^c_\lambda
\delta^{\rho\sigma}\Gamma^a_{bc}
-\frac{\partial B^a_\sigma}{\partial x^\mu}C^B_\nu C^D_\lambda
\delta^{\rho\sigma}\Gamma^A_{BD} \nonumber \\
 & & -B^b_\nu B^d_\mu B^e_\lambda\delta^{\rho\sigma}C^A_\sigma\Gamma^a_{bc}
\Gamma^c_{de}
-C^B_\nu C^E_\mu C^F_\lambda\delta^{\rho\sigma}B^a_\sigma\Gamma^A_{BD}
\Gamma^D_{EF}
-B^b_\nu\delta^{\rho\sigma}C^A_\sigma
\frac{\partial B^c_\lambda}{\partial x^\mu}\Gamma^a_{bc} \nonumber \\
 & & -C^B_\nu\delta^{\rho\sigma}B^a_\sigma\Gamma^A_{BD}
\frac{\partial C^D_\lambda}{\partial x^\mu}
+\frac{\partial C^A_\sigma}{\partial x^\nu}B^b_\mu B^c_\lambda
\delta^{\rho\sigma}\Gamma^a_{bc}
+\frac{\partial B^a_\sigma}{\partial x^\nu}C^B_\mu C^D_\lambda
\delta^{\rho\sigma}\Gamma^A_{BD})
\end{eqnarray}
\begin{eqnarray}
\Lambda' & = &
-\frac{\partial g_{aA}}{\partial x^\mu}B^b_\nu B^c_\lambda
\delta^{\rho\sigma}C^A_\sigma\Gamma^a_{bc}
-\frac{\partial g_{aA}}{\partial x^\mu}C^B_\nu C^D_\lambda
\delta^{\rho\sigma}B^a_\sigma\Gamma^A_{BD}
+\frac{1}{2}\frac{\partial g_{aA}}{\partial x^\nu}\delta^{\rho\sigma}
\frac{\partial(B^a_\sigma C^A_\lambda)}{\partial x^\mu} \nonumber \\
 & & +\frac{1}{2}\frac{\partial g_{aA}}{\partial x^\nu}\delta^{\rho\sigma}
\frac{\partial(C^A_\sigma B^a_\lambda)}{\partial x^\mu}
+\frac{1}{2}\frac{\partial g_{aA}}{\partial x^\lambda}
\delta^{\rho\sigma}\frac{\partial(B^a_\sigma C^A_\nu)}{\partial x^\mu}
+\frac{1}{2}\frac{\partial g_{aA}}{\partial x^\lambda}\delta^{\rho\sigma}
\frac{\partial(C^A_\sigma B^a_\nu)}{\partial x^\mu} \nonumber \\
 & & -\frac{1}{2}\frac{\partial g_{aA}}{\partial x^\sigma}
\delta^{\rho\sigma}\frac{\partial(B^a_\nu C^A_\lambda)}{\partial x^\mu}
-\frac{1}{2}\frac{\partial g_{aA}}{\partial x^\sigma}\delta^{\rho\sigma}
\frac{\partial(C^A_\nu B^a_\lambda)}{\partial x^\mu}
+\frac{\partial g_{aA}}{\partial x^\nu}B^b_\mu B^c_\lambda
\delta^{\rho\sigma}C^A_\sigma\Gamma^a_{bc} \nonumber \\
 & & +\frac{\partial g_{aA}}{\partial x^\nu}C^B_\mu C^D_\lambda
\delta^{\rho\sigma}B^a_\sigma\Gamma^A_{BD}
-\frac{1}{2}\frac{\partial g_{aA}}{\partial x^\mu}\delta^{\rho\sigma}
\frac{\partial(B^a_\sigma C^A_\lambda)}{\partial x^\nu}
-\frac{1}{2}\frac{\partial g_{aA}}{\partial x^\mu}\delta^{\rho\sigma}
\frac{\partial(C^A_\sigma B^a_\lambda)}{\partial x^\nu} \nonumber \\
 & & -\frac{1}{2}\frac{\partial g_{aA}}{\partial x^\lambda}
\delta^{\rho\sigma}\frac{\partial(B^a_\sigma C^A_\mu)}{\partial x^\nu}
-\frac{1}{2}\frac{\partial g_{aA}}{\partial x^\lambda}\delta^{\rho\sigma}
\frac{\partial(C^A_\sigma B^a_\mu)}{\partial x^\nu}
+\frac{1}{2}\frac{\partial g_{aA}}{\partial x^\sigma}\delta^{\rho\sigma}
\frac{\partial(B^a_\mu C^A_\lambda)}{\partial x^\nu} \nonumber \\
 & & +\frac{1}{2}\frac{\partial g_{aA}}{\partial x^\sigma}
\delta^{\rho\sigma}\frac{\partial(C^A_\mu B^a_\lambda)}{\partial x^\nu}
\end{eqnarray}
\begin{eqnarray}
\Phi' & = &
\frac{1}{2}\frac{\partial^2g_{aA}}{\partial x^\mu\partial x^\lambda}
\delta^{\rho\sigma}B^a_\sigma C^A_\nu
+\frac{1}{2}\frac{\partial^2g_{aA}}{\partial x^\mu\partial x^\lambda}
\delta^{\rho\sigma}C^A_\sigma B^a_\nu
-\frac{1}{2}\frac{\partial^2g_{aA}}{\partial x^\mu\partial x^\sigma}
\delta^{\rho\sigma}B^a_\nu C^A_\lambda \nonumber \\
 & & -\frac{1}{2}\frac{\partial^2g_{aA}}{\partial x^\mu\partial x^\sigma}
\delta^{\rho\sigma}C^A_\nu B^a_\lambda
-\frac{1}{2}\frac{\partial^2g_{aA}}{\partial x^\nu\partial x^\lambda}
\delta^{\rho\sigma}B^a_\sigma C^A_\mu
-\frac{1}{2}\frac{\partial^2g_{aA}}{\partial x^\nu\partial x^\lambda}
\delta^{\rho\sigma}C^A_\sigma B^a_\mu \nonumber \\
 & & +\frac{1}{2}\frac{\partial^2g_{aA}}{\partial x^\nu\partial x^\sigma}
\delta^{\rho\sigma}B^a_\mu C^A_\lambda
+\frac{1}{2}\frac{\partial^2g_{aA}}{\partial x^\nu\partial x^\sigma}
\delta^{\rho\sigma}C^A_\mu B^a_\lambda
\end{eqnarray}
are the contributions of $g_{aA}$, $\partial g_{aA}$ and
$\partial^2g_{aA}$, respectively. Correspondingly, $R^{\ \ \ d}_{abc}$ in
(69) becomes
\begin{equation}
R^{\ \ \ d}_{abc}=D_a(g^{de}B^\mu_e\delta_{\mu\nu})D_bB^\nu_c
-D_b(g^{de}B^\mu_e\delta_{\mu\nu})D_aB^\nu_c
\end{equation}
in which
\begin{equation}
D_a(g^{de}B^\mu_e\delta_{\mu\nu})
=\frac{\partial(g^{de}B^\mu_e\delta_{\mu\nu})}{\partial u^a}
+\Gamma^d_{ab}g^{be}B^\mu_e\delta_{\mu\nu}
\end{equation}
\begin{equation}
D_bB^\nu_c=\frac{\partial B^\nu_c}{\partial u^b}-\Gamma^a_{bc}B^\nu_a
\end{equation}
are the covariant derivatives of $g^{de}B^\mu_e\delta_{\mu\nu}$ and
$B^\nu_c$ in the present condition. From (77) $R^{\ \ \ d}_{abc}$ is further
read as
\begin{eqnarray}
R^{\ \ \ d}_{abc} & = &
\frac{\partial(g^{di}B^\lambda_i\delta_{\lambda\rho})}{\partial u^a}
(C^\rho_A-B^\rho_fg^{fe}g_{eA})C^A_\nu\frac{\partial B^\nu_c}{\partial u^b}
\nonumber \\
 & & -\frac{\partial(g^{di}B^\lambda_i\delta_{\lambda\rho})}{\partial u^b}
(C^\rho_A-B^\rho_fg^{fe}g_{eA})C^A_\nu\frac{\partial B^\nu_c}{\partial u^a}.
\end{eqnarray}
Since $C^\mu_A$ has nothing to do with $u^a$, we obtain
\begin{eqnarray}
R^{\ \ \ d}_{abc} & = &
\frac{\partial(g^{de}g_{eA})}{\partial u^a}C^A_\mu
\frac{\partial B^\mu_c}{\partial u^b}
-\frac{\partial(g^{de}B^\mu_e\delta_{\mu\nu})}{\partial u^a}B^\nu_fg^{fg}
g_{gA}C^A_\lambda\frac{\partial B^\lambda_c}{\partial u^b} \nonumber \\
 & & -\frac{\partial(g^{de}g_{eA})}{\partial u^b}C^A_\mu
\frac{\partial B^\mu_c}{\partial u^a}
+\frac{\partial(g^{de}B^\mu_e\delta_{\mu\nu})}{\partial u^b}B^\nu_fg^{fg}
g_{gA}C^A_\lambda\frac{\partial B^\lambda_c}{\partial u^a} \nonumber \\
 & = & D_a(g^{de}g_{eA})C^A_\mu\frac{\partial B^\mu_c}{\partial u^b}
-D_b(g^{de}g_{eA})C^A_\mu\frac{\partial B^\mu_c}{\partial u^a}.
\end{eqnarray}
We see that when $g_{aA}=0$, $R^{\ \ \ d}_{abc}=0$ and when 
$g_{aA}\neq 0$, $R^{\ \ \ d}_{abc}$ may not vanish. What does this mean?
From (34) we know that when $g_{aA}=0$, the global manifold $X$ is the
direct product of $M$ and $N$. Under this condition, (88) tells us when $X$
is Euclidean, both the two submanifolds $M$ and $N$ are Euclidean, too, i.e.
$R^k=R^m\times R^n$. This is the trivial case either in mathematic or
in physics. However, when $g_{aA}\neq 0$, $X$ can not be expressed as the
direct product of $M$ and $N$. Even when $X$ is Euclidean, $M$ and $N$ are
not Euclidean but affect each other through $g_{aA}$. In this sense, we say
$g_{aA}$ is the bridge between $M$ and $N$. After the symmetrization, the
Ricci tensor of $M$ is
\begin{eqnarray}
R_{bc} & = & -\frac{1}{2}(g^{af}B^\nu_f\delta_{\nu\lambda}
\frac{\partial(B^\lambda_dg^{de}g_{eA})}{\partial u^b}C^A_\mu
\frac{\partial B^\mu_c}{\partial u^a}
+g^{af}B^\nu_f\delta_{\nu\lambda}
\frac{\partial(B^\lambda_dg^{de}g_{eA})}{\partial u^c}C^A_\mu
\frac{\partial B^\mu_b}{\partial u^a}) \nonumber \\
 & & +g^{af}B^\nu_f\delta_{\nu\lambda}
\frac{\partial(B^\lambda_dg^{de}g_{eA})}{\partial u^a}C^A_\mu
\frac{\partial B^\mu_c}{\partial u^b}
\end{eqnarray}
which gives the scalar curvature
\begin{equation}
R=-B^\sigma_ag^{af}B^\nu_f\delta_{\nu\lambda}
(\frac{\partial(B^\lambda_dg^{de}g_{eA})}{\partial x^\sigma}
\frac{\partial C^A_\mu}{\partial x^\rho}
-\frac{\partial(B^\lambda_dg^{de}g_{eA})}{\partial x^\rho}
\frac{\partial C^A_\mu}{\partial x^\sigma})B^\rho_bg^{bc}B^\mu_c.
\end{equation}
Then one has
\begin{eqnarray}
R_{ij}-\frac{1}{2}g_{ij}R & = &
-\frac{1}{2}B^\sigma_ag^{af}B^\nu_f\delta_{\nu\lambda}
(\frac{\partial(B^\lambda_dg^{de}g_{eA})}{\partial x^\sigma}
\frac{\partial C^A_\mu}{\partial x^\rho}
-\frac{\partial(B^\lambda_dg^{de}g_{eA})}{\partial x^\rho}
\frac{\partial C^A_\mu}{\partial x^\sigma}) \nonumber \\
 & & B^\rho_bB^\mu_c(\delta^b_i\delta^c_j+\delta^b_j\delta^c_i-g^{bc}g_{ij}).
\end{eqnarray}
Corresponding to the Einstein equation of $M$
\begin{equation}
R_{ij}-\frac{1}{2}g_{ij}R=-8\pi GT_{ij},
\end{equation}
the energy-momentum tensor of matter field in $M$ is determined by
\begin{eqnarray}
T_{ij} & = &
\frac{1}{16\pi G}B^\sigma_ag^{af}B^\nu_f\delta_{\nu\lambda}
(\frac{\partial(B^\lambda_dg^{de}g_{eA})}{\partial x^\sigma}
\frac{\partial C^A_\mu}{\partial x^\rho}
-\frac{\partial(B^\lambda_dg^{de}g_{eA})}{\partial x^\rho}
\frac{\partial C^A_\mu}{\partial x^\sigma}) \nonumber \\
 & & B^\rho_bB^\mu_c(\delta^b_i\delta^c_j+\delta^b_j\delta^c_i-g^{bc}g_{ij})
\end{eqnarray}
in which $G$ is the Newton constant. Here one can see that, in order to
obtain the nonzero energy-momentum tensor $T_{ij}$ of matter field, there
must be the contributions of the bridge $g_{aA}$ and the transversal
submanifold $N$. So we conclude that the other transversal submanifold $N$
is closely related to the matter fields of the submanifold $M$. This
conclusion is the significant generalization to the
conventional submanifold theory of Riemann geometry and it is in agreement
with the Kaluza-Klein theory, of which the high-dimensional coordinates of
space-time are linked to the gauge fields through the so-called dimensional
reduction.

\vskip 48pt
\noindent
{\bf V. CONCLUSIONS}

\vskip 24pt
\indent In this paper, by replacing the normal vector of a submanifold with
the tangent vector of another submanifold, the total relations of the metric
tensors, Christoffel symbols and curvature tensors of two transversal
submanifolds and global manifold are discussed. The corresponding formulas are
the generalizations of the conventional submanifold theory of Riemann
geometry. When the inner product of the two tangent vectors vanishes, all of
the above consequences will return to the conventional submanifold theory and
some corollaries will give the most important second fundamental form and
Gauss-Codazzi equation. As a special case, the global manifold is Euclidean is
considered. It is pointed out that, in order to maintain the property of the
given global manifold, the metric tensors, Christoffel symbols and curvature
tensors of the two transversal submanifolds must satisfy some conditions.
Furthermore, the above inner product plays the role of bridge through which
the two transversal submanifolds affect each other and one submanifold is
closely related to the matter fields of another submanifold. This conclusion
is the significant generalization to the conventional submanifold theory of
Riemann geometry and it is in agreement with the Kaluza-Klein theory, of which
the high-dimensional coordinates of space-time are linked to the gauge fields
through the so-called dimensional reduction. The consequences obtained in this
paper can be applied to generalize the models of direct product of manifolds
in string and D-brane theories to the more general cases --- nondirect product
manifolds. The applications of this paper will be studied in our later work.

\vskip 48pt
\noindent
{\bf ACKNOWLEDGEMENTS}

\vskip 24pt
\indent The first author thanks for the discussions with Prof. Shi-Xiang Feng.
This work was supported by the National Nature Science Foundation of China
under Grant No. 19805004.

\vskip 48pt
\noindent
\mbox{\hspace{4pt}}$^1$J. Erlich, A. Naqvi and L. Randall, Phys. Rev. D
{\bf 58}, 046002 (1998).\\
\noindent
\mbox{\hspace{4pt}}$^2$S. Hyun, Y. Kiem and H. Shin, Phys. Rev. D {\bf 57},
4856 (1998).\\
\noindent
\mbox{\hspace{4pt}}$^3$C.S. Chu, P.S. Howe and E. Sezgin, Phys. Lett. B
{\bf 428}, 59 (1998).\\
\noindent
\mbox{\hspace{4pt}}$^4$I. Kishimoto and N. Sasakura, Phys. Lett. B {\bf 432},
305 (1998).\\
\noindent
\mbox{\hspace{4pt}}$^5$S.W. Hawking and M.M. Taylor-Robinson, Phys. Rev. D
{\bf 58}, 025006 (1998).\\
\noindent
\mbox{\hspace{4pt}}$^6$S. Ferrara, A. Kehagias, H. Partouche and A. Zaffaroni,
Phys. Lett. B {\bf 431}, 42 (1998).\\
\noindent
\mbox{\hspace{4pt}}$^7$G.H. Yang and Y.S. Duan, Mod. Phys. Lett. A {\bf 13},
745 (1998).\\
\noindent
\mbox{\hspace{4pt}}$^8$G.H. Yang, Mod. Phys. Lett. A {\bf 13}, 2123 (1998).\\
\noindent
\mbox{\hspace{4pt}}$^9$L. Eisenhart, {\it Riemannian Geometry} (Princeton
University Press, Princeton, NJ, 1964).
\end{document}